\documentclass[sigconf]{aamas}

\usepackage{balance}
\usepackage{graphicx}
\usepackage{float}
\usepackage{placeins}
\usepackage[inline]{enumitem}
\usepackage{fancyvrb}
\usepackage{caption}

\makeatletter

\gdef\@copyrightpermission{
  \begin{minipage}{0.2\columnwidth}
   \href{https://creativecommons.org/licenses/by/4.0/}{\includegraphics[width=0.90\textwidth]{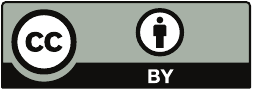}}
  \end{minipage}\hfill
  \begin{minipage}{0.8\columnwidth}
   \href{https://creativecommons.org/licenses/by/4.0/}{This work is licensed under a Creative Commons Attribution International 4.0 License.}
  \end{minipage}
  \vspace{5pt}
}
\makeatother

\title[MOASEI Competition AAMAS'2025 Technical Report]{Inaugural MOASEI Competition at AAMAS’2025: A Technical Report}

\author{Ceferino Patino}
\affiliation{
  \institution{University of Nebraska - Lincoln}
  \city{Lincoln}
  \state{Nebraska}
  \country{United States}}
\email{cpatino2@unl.edu}

\author{Tyler J. Billings}
\affiliation{
  \institution{University of Nebraska - Lincoln}
  \city{Lincoln}
  \state{Nebraska}
  \country{United States}}
\email{tbillings4@unl.edu}

\author{Alireza Saleh Abadi}
\affiliation{
  \institution{University of Nebraska - Lincoln}
  \city{Lincoln}
  \state{Nebraska}
  \country{United States}}
\email{asalehabadi2@unl.edu}

\author{Daniel Redder}
\affiliation{
  \institution{University of Georgia - Athens}
  \city{Athens}
  \state{Georgia}
  \country{United States}}
\email{daniel.redder@uga.edu}

\author{Adam Eck}
\affiliation{
  \institution{Oberlin College}
  \city{Oberlin}
  \state{Ohio}
  \country{United States}}
\email{aeck@oberlin.edu}

\author{Prashant Doshi}
\affiliation{
  \institution{University of Georgia - Athens}
  \city{Athens}
  \state{Georgia}
  \country{United States}}
\email{pdoshi@uga.edu}

\author{Leen-Kiat Soh}
\affiliation{
  \institution{University of Nebraska - Lincoln}
  \city{Lincoln}
  \state{Nebraska}
  \country{United States}}
\email{lksoh@unl.edu}

\setcopyright{rightsretained}
\acmConference[]{}
\copyrightyear{2025}
\acmYear{}
\acmDOI{}
\acmPrice{}
\acmISBN{}

\begin{abstract} 
We present the Methods for Open Agent Systems Evaluation Initiative (MOASEI) Competition, a multi-agent AI benchmarking event designed to evaluate decision-making under open-world conditions. Built on the free-range-zoo environment suite, MOASEI introduced dynamic, partially observable domains with agent and task openness--settings where entities may appear, disappear, or change behavior over time. The 2025 competition featured three tracks--Wildfire, Rideshare, and Cybersecurity--each highlighting distinct dimensions of openness and coordination complexity. Eleven teams from international institutions participated, with four of those teams submitting diverse solutions including graph neural networks, convolutional architectures, predictive modeling, and large language model--driven meta--optimization. Evaluation metrics centered on expected utility, robustness to perturbations, and responsiveness to environmental change. The results reveal promising strategies for generalization and adaptation in open environments, offering both empirical insight and infrastructure for future research. This report details the competition’s design, findings, and contributions to the open-agent systems research community. 
\end{abstract}


\newcommand{\BibTeX}{\rm B\kern-.05em{\sc i\kern-.025em b}\kern-.08em\TeX}

\newcommand{\pinoop}{$\pi_{\texttt{noop}}$}
\newcommand{\pirandom}{$\pi_{\texttt{random}}$}

\newcommand{\pilargest}{$\pi_{\texttt{largest}}$}
\newcommand{\pismallest}{$\pi_{\texttt{smallest}}$}

\newcommand{\pipatched}{$\pi_{\texttt{patched}}$}
\newcommand{\piexploited}{$\pi_{\texttt{exploited}}$}
\begin{document}

\pagestyle{fancy}
\fancyhead{}

\maketitle

\section{Introduction}

To promote research on open-world decision-making and foster community engagement, we organized the Methods for Open Agent Systems Evaluation Initiative (MOASEI) Competition, an international benchmarking event focused on evaluating multi-agent AI systems under conditions of agent and task openness. The competition was built on top of the free-range-zoo environment suite and served as a platform for exploring how AI agents adapt to dynamic and evolving environments. The inaugural competition was announced in March 2025 and concluded on May 19th, 2025 at the AAMAS 2025 conference in Detroit, MI, USA.

In the following sections, we detail the structure, participation, and outcomes of the MOASEI Competition. We describe the competition tracks and evaluation metrics, summarize participant engagement and submitted approaches, and highlight key empirical findings from the event. Our goal is to present MOASEI not only as a proof-of-concept for open-system benchmarking, but also as a foundation for future work on evaluating and improving AI systems operating under real-world conditions of uncertainty, dynamism, and limited control.

\section{Background}

Real-world multi-agent systems often operate under conditions of uncertainty, limited information, and dynamic task requirements. To model such complexity, we adopt the framework of \emph{partially observable stochastic games} (POSGs), which generalize Markov decision processes to multi-agent settings with partial observability. In this section, we first define the general POSG formulation used throughout the competition. We then provide domain-specific formalizations for each of the three open environments--wildfire suppression, cybersecurity defense, and dynamic ridesharing--highlighting the unique structural and observational challenges posed by each.

\subsection{Partially Observable Stochastic Games}

We utilized three separate open environments for the competition~\cite{eck2023_decision}. We formulate each domain as a partially observable stochastic game (POSG), where each environment is defined by the tuple $\{\mathcal{S}, \mathcal{A}, \mathcal{T}, \mathcal{R}, \mathcal{Z}, \mathcal{O}\}$. Here, $\mathcal{S}$, $\mathcal{A}$, and $\mathcal{Z}$ represent the state, action, and observation spaces, respectively. The transition function $\mathcal{T}: \mathcal{S} \times \mathcal{A} \rightarrow \Delta(\mathcal{S})$ defines the probability distribution over next states given the current state and joint actions $\mathbf{a}$. The reward function $\mathcal{R}: \mathcal{S} \times \mathcal{A} \times \mathcal{S} \rightarrow \mathbb{R}^n$ maps state transitions and joint actions to a vector of agent-specific rewards. The observation function $\mathcal{O}: \mathcal{S} \times \mathcal{A} \rightarrow \Delta(\mathcal{Z}^n)$ defines the distribution over each agent’s observations given the state and joint action.

\subsection{Formal Definition for Wildfire Suppression}

The wildfire suppression environment models a decentralized team of agents combating spatially spreading fires. The state $\mathcal{S}$ consists of a set of agents characterized by internal properties such as position $(y, x)$, power, and suppressant capacity, alongside a set of fires defined by their location, the number of agents required to suppress them, and their size, including metrics like distance from burnout or putout status. Each agent’s action space $\mathcal{A}$ includes commands to fight one of the adjacent fires $\{\texttt{fight}_0, \dots, \texttt{fight}_n\}$, or to perform no operation (\texttt{noop}). The transition function $\mathcal{T}$ encompasses fire spread dynamics, which propagate existing fires to neighboring areas and capture task openness, as well as suppressant consumption and replenishment mechanics, which govern agent resource availability and model agent openness. The reward function $\mathcal{R}$ penalizes the team by $-1$ for each fire that burns out and rewards them by $2^{\text{agents\_required}}$ for every fire successfully extinguished in a timestep. The observation space $\mathcal{Z}$ includes all internal agent properties, the positions of other agents, and the attributes of all active fires. Finally, the observation function $\mathcal{O}$ provides each agent with a partial, deterministic view of the environment as specified by $\mathcal{Z}$. Additionally, this environment contains a termination criterion, where the simulation is immediately terminated if a state is ever reached where there are no fires in the environment.

\subsection{Formal Definition for Cybersecurity Defense}

The cybersecurity defense environment models interactions between defender and attacker agents across a network of interconnected subnetworks. The state $\mathcal{S}$ includes sets of defenders and attackers, each characterized by internal properties such as location, power, and presence status. Additionally, the state includes subnetworks and their properties, including their current state and outgoing connections. The action space $\mathcal{A}$ for each agent comprises possible moves to different nodes ($\{\texttt{move}_0, \dots, \texttt{move}_n\}$), no operation (\texttt{noop}), patching the current node (\texttt{patch}), and monitoring all nodes (\texttt{monitor}). The transition function $\mathcal{T}$ governs the presence dynamics, dictating when attacker or defender agents lose or regain connection to the environment, and models the state changes of subnetworks based on the relative exploit power of attackers versus patching power of defenders at each node. The reward function $\mathcal{R}$ aggregates values from the set $\{4, 0, -2, -4, -8\}$ scaled by $2^{\text{outgoing connections}}$ for each node, with this reward received at every simulation step. The observation space $\mathcal{Z}$ includes each agent’s internal properties and the power levels of all other agents; however, the properties of subnetworks are only observed if explicitly monitored. Finally, the observation function $\mathcal{O}$ provides partial, deterministic views of the environment as defined by $\mathcal{Z}$.

\subsection{Formal Definition for Dynamic Ridesharing}

The dynamic ridesharing environment models a set of driver agents and passenger tasks distributed spatially. The state $\mathcal{S}$ consists of drivers characterized by internal properties such as position $(y, x)$, the number of accepted passengers, and the number of currently riding passengers. It also includes passenger tasks with properties including position $(y, x)$, destination coordinates $(\text{x}_{dest}, \text{y}_{dest})$, fare, state (unaccepted, accepted, picked up), agent association, and timestamps for when the task entered the system and was accepted or picked up. The action space $\mathcal{A}$ is defined by the set $\{(\texttt{accept}|\texttt{pick}|\texttt{drop})_0, \dots, (\texttt{accept}|\texttt{pick}|\texttt{drop})_n\}$, which represents the actions to accept, pick up, or drop off passengers, respectively, for indexed tasks where the available action depends on the state of each task, as well as a no-operation action (\texttt{noop}). The transition function $\mathcal{T}$ includes passenger entry dynamics defining when new tasks are added, passenger state transitions progressing tasks through $\{unaccepted, accepted, picked-up, dropped-off\}$ states, and driver movement transitions which move drivers and their currently riding passengers. The reward function $\mathcal{R}$ is defined by fares from successfully completed trips, typically the Manhattan distance between start and destination, and includes global penalties for waiting passengers. The observation space $\mathcal{Z}$ contains each driver’s internal properties (position, number of accepted and riding passengers), the same properties for other drivers, and task properties for unaccepted (global) tasks and those directly associated with the driver agent (accepted or picked up), including location, destination, assigned driver agent, and timing information. Finally, the observation function $\mathcal{O}$ provides each driver agent with a partial deterministic view derived from $\mathcal{Z}$ based on the true environment state.

\section{Overview of the MOASEI Competition}

MOASEI directly supported the project's goal of evaluating how openness affects the performance and generalization of AI systems in complex, multi-agent environments. It provided a practical testbed for comparing strategies and encouraged participants to confront real-world-inspired challenges that traditional closed-environment competitions often overlook. The competition also served as a dissemination channel for the free-range-zoo framework and helped establish a shared benchmark for future research in open-world AI.

\subsection{Design of Competition Tracks and Metrics}

The competition featured \textbf{three tracks} using different open system domains \cite{eck2023_decision}, each highlighting different dimensions of openness:

\begin{enumerate}
	\item\textbf{Wildfire} incorporated both agent and task openness, requiring participants to reason about shifting agent availability and dynamically appearing tasks in a high-stakes, coordination-heavy setting.
	\item\textbf{Rideshare} focused on task openness, where agents prioritize delivering passengers who enter unpredictably over time..
	\item\textbf{Cybersecurity} emphasized agent openness, where defenders entered and exited the environment unpredictably, simulating the challenge of managing rotating personnel or automated defense systems.
\end{enumerate}

Each track was evaluated using a consistent set of performance metrics, including expected utility of the subject policy, robustness to modulations in the level of openness in the environment, and responsiveness to changes in agent / task availability. Each domain was also evaluated with additional domain specific metrics, such as the efficiency of the policy and preferences towards types of actions. Example environment configurations and baseline policies were released alongside detailed documentation to support participant onboarding.

\subsection{Recruitment and Participation}

We conducted outreach to academic institutions, AI research labs, and online communities to encourage participation. As a result, eleven teams from international universities and research organizations registered. Notably, four of the registered teams submitted solutions for the \textbf{wildfire} and \textbf{cybersecurity} tracks, while the \textbf{rideshare} track saw no participation in this cycle. Participant engagement demonstrated significant interest in open-system challenges and validated the accessibility of the provided environments and documentation. We also organized two separate virtual "office hours" in April 2025. The office hours gave competition teams the opportunity to voice questions questions and resolve issues related to our codebase, domains, and other aspects of the competition.

Throughout the competition, the organizers maintained active communication with participants to ensure clarity and engagement. Regular reminder emails were sent to highlight upcoming deadlines and key milestones. The organizers also responded to participant inquiries regarding the codebase, evaluation methods, and overall competition logistics. In addition to direct communication, a dedicated competition website \cite{moasei2025_website} was maintained as a central hub for essential information. This included the problem background, competition schedule, descriptions of the various tracks, evaluation guidelines, and official rules. The website was regularly updated with new details, announcements, and links to virtual office hours to support ongoing participant engagement. To accommodate participants and team members unable to attend the conference in person, the competition results was made available virtually to them, ensuring broad accessibility.

\subsection{Infrastructure, Evaluation, and Results Analysis}

We implemented an automated evaluation pipeline to ensure reproducibility and fairness across teams. Submissions were tested on unseen instances with varying degrees of openness, using fixed seeds for statistical comparison. We compiled the results into leaderboards and analyzed to identify the strengths and weaknesses of different policy approaches, their sensitivity to various types of openness, and emergent strategies for task allocation and coordination.

For a detailed discussion of the evaluation configurations, domain-specific metrics, and planned enhancements to the evaluation infrastructure, please refer to Section~\ref{sec:eval_methods_and_results} on Evaluation Methodology and Results Analysis.

The free-range-zoo codebase \cite{moasei2025_repository} forms the primary foundation of the competition and is publicly available for download. It provides complete implementations of the three benchmark domains --- wildfire, cybersecurity, and rideshare --- with a common interface and consistent abstractions for agents, tasks, and environment dynamics. Each domain is implemented as a standalone module, enabling competitors to interact with them independently for evaluation and experimentation.

To facilitate participant engagement and reproducibility, the repository includes a set of starter scripts for evaluation. These scripts allow teams to run trained agents in controlled scenarios and generate metrics compatible with the official leaderboard evaluation protocol. An interactive rendering script is also included, enabling teams to visually inspect environment rollouts in real-time or replay saved runs. The renderer supports key visual overlays, including agent positions, task markers, and temporal progression, making it a useful tool for debugging and qualitative policy analysis.

The repository also provides example training configurations tailored to each domain. These configuration files specify key environment parameters and are intended to serve as a baseline for teams developing their own training pipelines. While no learning algorithms are bundled with the codebase, the provided configurations and environment wrappers are compatible with most off-the-shelf reinforcement learning libraries.

Extensive documentation accompanies the release. A full user manual is included with instructions for running environments, using the evaluation and rendering tools, and evaluating custom policies. An additional developer reference describes the environment interface and the codebase structure, helping teams extend or wrap the environment to restructure information as needed.

\begin{figure}[ht]
\centering
\begin{minipage}{0.9\linewidth}
\begin{Verbatim}
| <environment> # <environment implementation>
+-- configs     # benchmark configurations
+-- baselines   # baseline policies
+-- env         # environment definitions
    +-- spaces      # action / observation spaces
    +-- structures  # environment config and state
    +-- transitions # transition functions
    +-- <environment>.py # environment definition
\end{Verbatim}
\captionof{figure}{Repository structure provided to competitors for each of the domains}
\label{fig:free_range_zoo_structure}
\end{minipage}
\end{figure}

The repository is organized for clarity and extensibility. The top-level directory includes subfolders for domain environments, training and evaluation scripts, configuration templates, and documentation. Figure~\ref{fig:free_range_zoo_structure} summarizes the structure of each of the domains. Together, these materials are designed to support both competition participation and broader research in open multi-agent systems.

\subsection{Outreach}

Collaborators held a technical tutorial at AAMAS’2025 on May 19, 2025, from 8:30 AM to 12:00 noon. The tutorial covered a range of topics with hands-on group activities. The topics included an introduction to open agent systems (OASYS) (45 minutes), multiagent reasoning in OASYS (50 minutes), emerging and future research in OASYS (45 minutes), and the MOASEI competition and community discussion (20 minutes), with a 45-minute mandatory coffee break. There were between 15-21 attendees at our tutorial. The attendees were actively engaged in our tutorial and discussions. 

\section{Evaluation Methodology and Results Analysis}
\label{sec:eval_methods_and_results}

To ensure rigorous and reproducible assessment, each team submission was evaluated using an automated pipeline, executing $n = 256$ independent simulation runs per policy across three held-out evaluation configurations. These evaluation scenarios were intentionally disjoint from training configurations to measure policy generalization under novel conditions. Statistical signficance is determined with $p < 0.05$.

We utilize average cumulative episodic reward of multiple runs as the primary performance metric across all tracks. In addition, each domain included a set of specialized metrics tailored to its specific dynamics and challenges. For instance, in the wildfire domain we monitored fire suppression efficiency and agent engagement; in cybersecurity, defensive activity distribution and temporal performance trends. These metrics are described in detail in the subsections that follow.

To support fair comparison, all evaluations used fixed random seeds and consistent scenario generation procedures. Aggregated results were presented in leaderboards and analyzed to uncover behavioral trends, level of openness, and coordination strategies. This structured evaluation framework not only benchmarked performance but also revealed actionable insights for improving the design of both participant policies and our open-environment testbeds.

\subsection{Wildfire Track Results}

The wildfire track featured both agent and task openness, requiring policies to adapt to dynamic task appearances and fluctuating agent availability in a spatially distributed grid environment.

Evaluation metrics for this domain included:
\begin{itemize}
	\item\textbf{Fires extinguished:} Total number of fires successfully put out by agents.
	\item\textbf{Burnouts:} Number of fires that burned out without intervention, often indicating poor coordination or insufficient agent coverage.
	\item\textbf{Simulation length:} The total number of steps until all fires have either burned out or been extinguished, reflecting how long agents maintained control of the environment.
  \item\textbf{Action distribution:} Proportions of agent actions (e.g., fighting fires or idling) to identify engagement levels and strategic behavior.
\end{itemize}

Table~\ref{tab:wildfire_policy_rewards} shows the average episodic rewards
across submitted and baseline policies for the wildfire domain.  There were
four baseline policies: 
\begin{enumerate*}[label=(\arabic*)]
  \item\pinoop, which remains idle throughout the simulation;
  \item\pirandom, which uniformly samples valid random actions;
  \item\pismallest, which chooses to fight the smallest fire; and
  \item\pilargest, which chooses to fight the largest fire.
\end{enumerate*}
The three submissions from the competition were BIT Student (using an LLM-based
approach), Markov Mayhem (using a GNN-based approach), and University of Tehran
(using a CNN-based approach.  

\begin{figure}[htbp]
	\centering
	\includegraphics[width=\columnwidth, trim=15 15 15 15, clip]{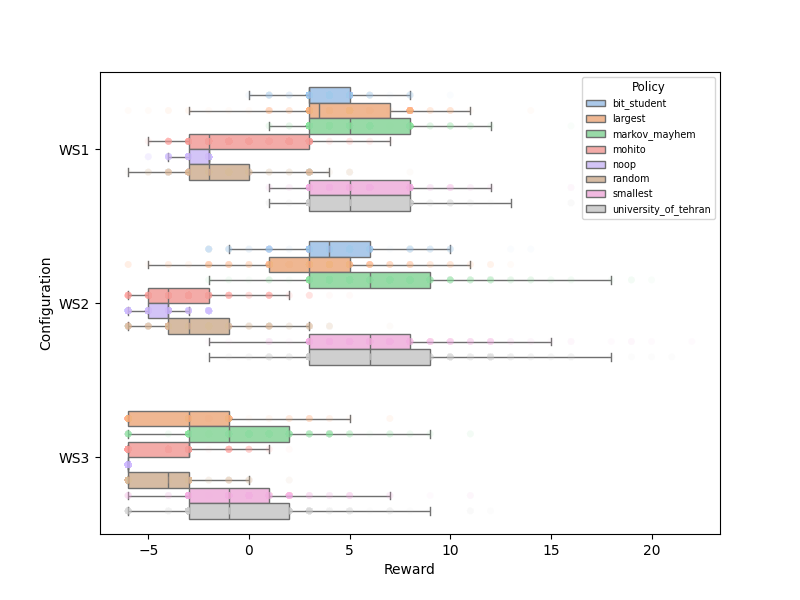}
	\caption{Wildfire expected rewards on a per-policy basis}
	\label{fig:wildfire_reward_plot}
\end{figure}

\begin{figure}[H]
	\centering
	\includegraphics[width=\columnwidth]{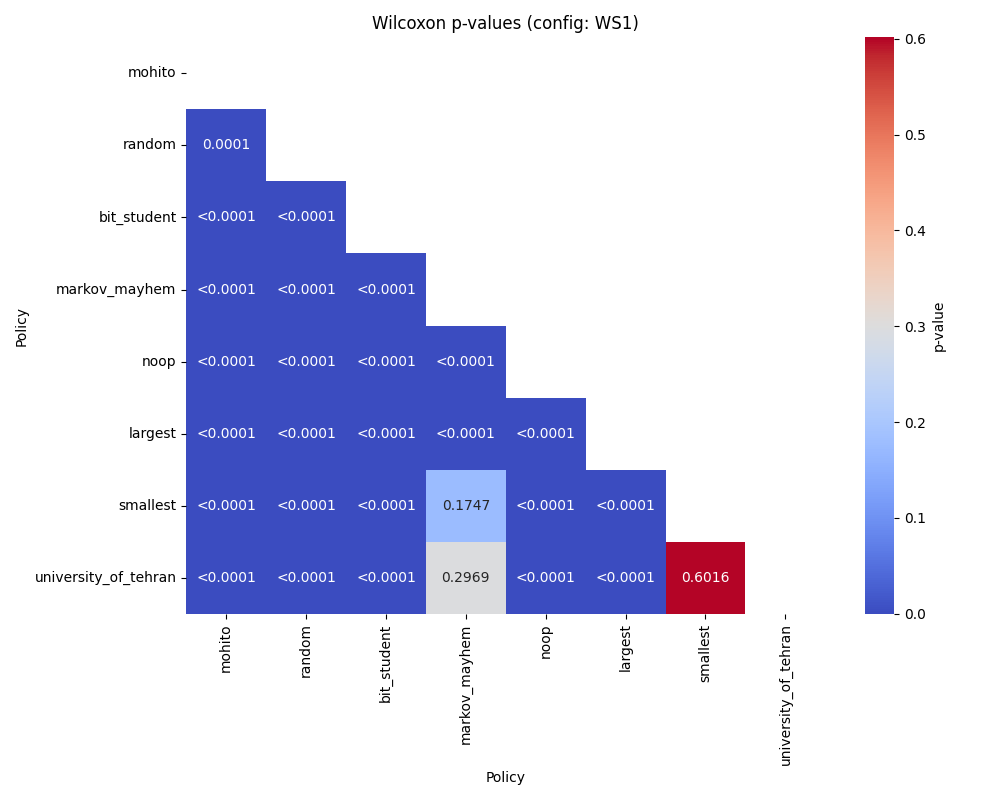}
	\caption{Wilcoxon-signed-rank comparisons for WS1}
	\label{fig:wildfire_reward_wilcoxon_ws1}
\end{figure}

\begin{figure}[H]
	\centering
	\includegraphics[width=\columnwidth]{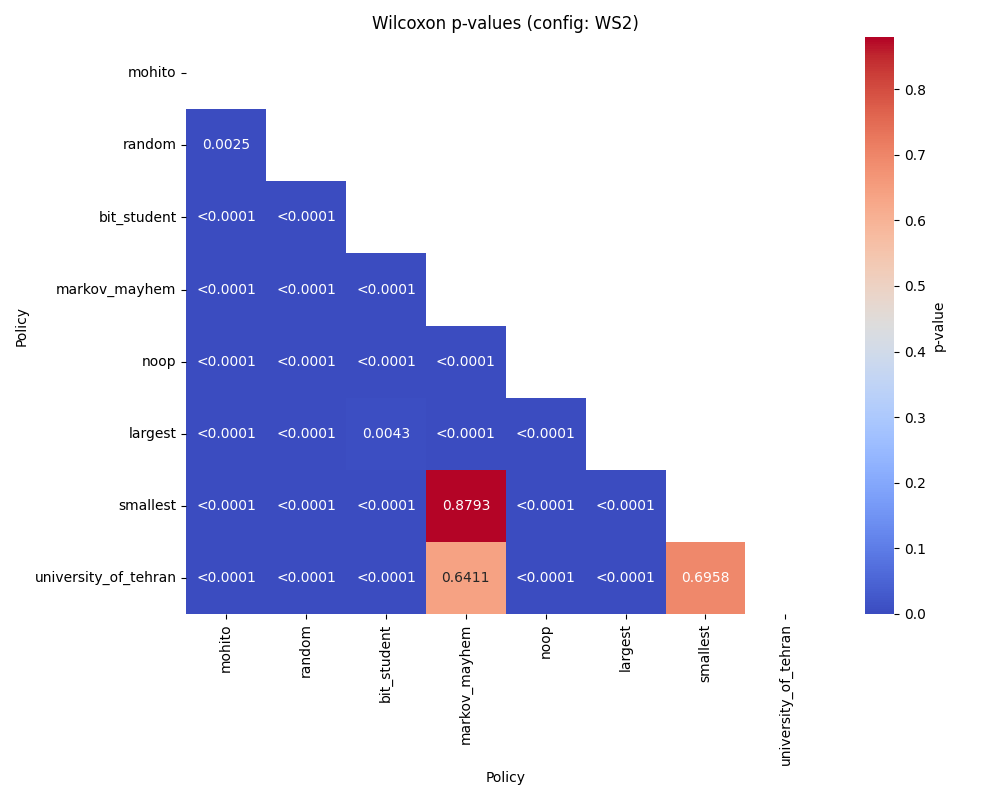}
	\caption{Wilcoxon-signed-rank comparisons for WS2}
	\label{fig:wildfire_reward_wilcoxon_ws2}
\end{figure}

\begin{figure}[H]
	\centering
	\includegraphics[width=\columnwidth]{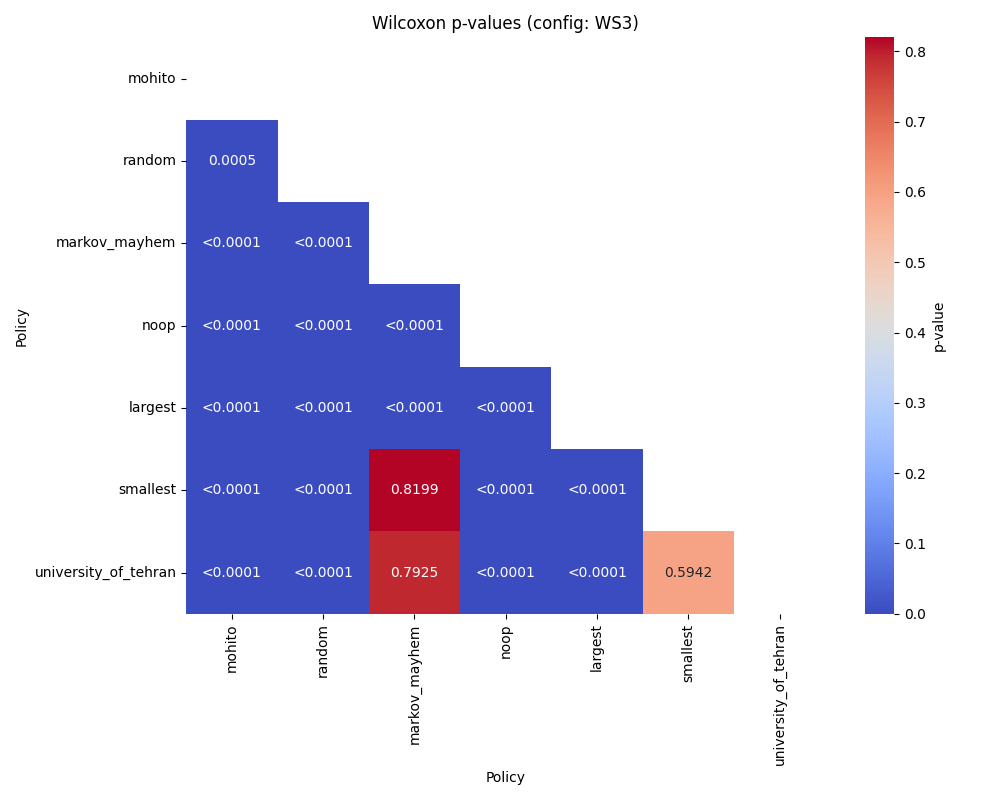}
	\caption{Wilcoxon-signed-rank comparisons for WS3}
	\label{fig:wildfire_reward_wilcoxon_ws3}
\end{figure}

Figure~\ref{fig:wildfire_reward_plot} and Table~\ref{tab:wildfire_policy_rewards} shows the expected rewards on a per-policy basis for the different approaches in the wildfire domain. Figures \ref{fig:wildfire_reward_wilcoxon_ws1}, \ref{fig:wildfire_reward_wilcoxon_ws2}, and \ref{fig:wildfire_reward_wilcoxon_ws3} show the Wilcoxon-signed-rank comparisons for the different approaches' performance to determine statistically significant differences in rewards. We observe that Markov Mayhem and University of Tehran achieved the highest cumulative episodic rewards out of all the submissions, however did not have statistically signficiant differences between each other and smallest.

Tables \ref{tab:wildfire_policy_fights} and \ref{tab:wildfire_policy_noop} show the proportions of choosing the fight a fire (Fight) action and choosing to do nothing (NoOp) for each of the approaches. We observe that the Bit Student submission tends towards performing significantly less fight actions in WS1 and WS2, while submissions by Markov Mayhem and University of Tehran tend towards fight-heavy policies.

\begin{figure}[H]
	\centering
	\includegraphics[width=\columnwidth, trim=15 15 15 15, clip]{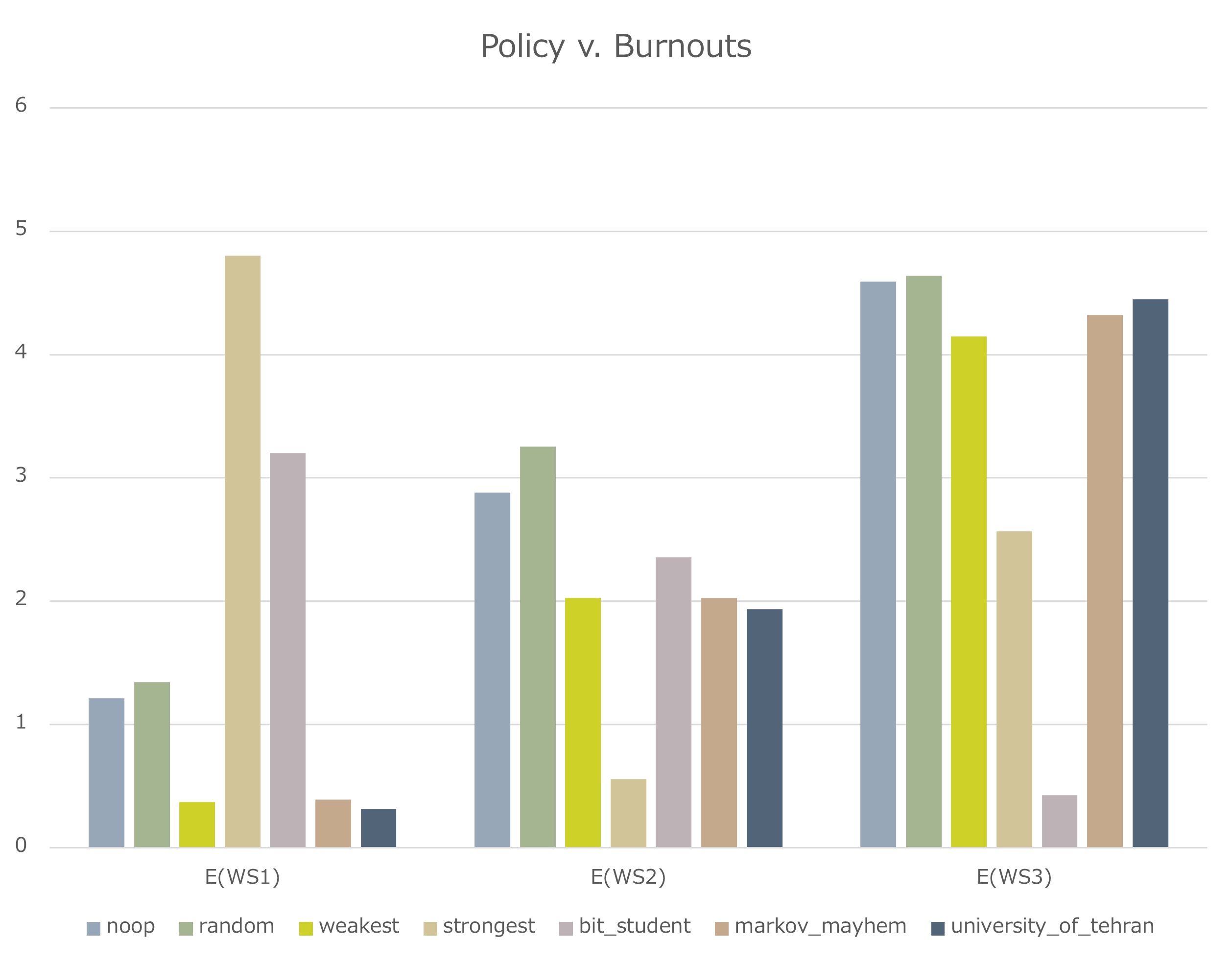}
	\caption{Burnouts sustained by each policy in each test configuration}
	\label{fig:wildfire_policy_burnouts}
\end{figure}

\begin{figure}[H]
	\centering
	\includegraphics[width=\columnwidth, trim=15 15 15 15, clip]{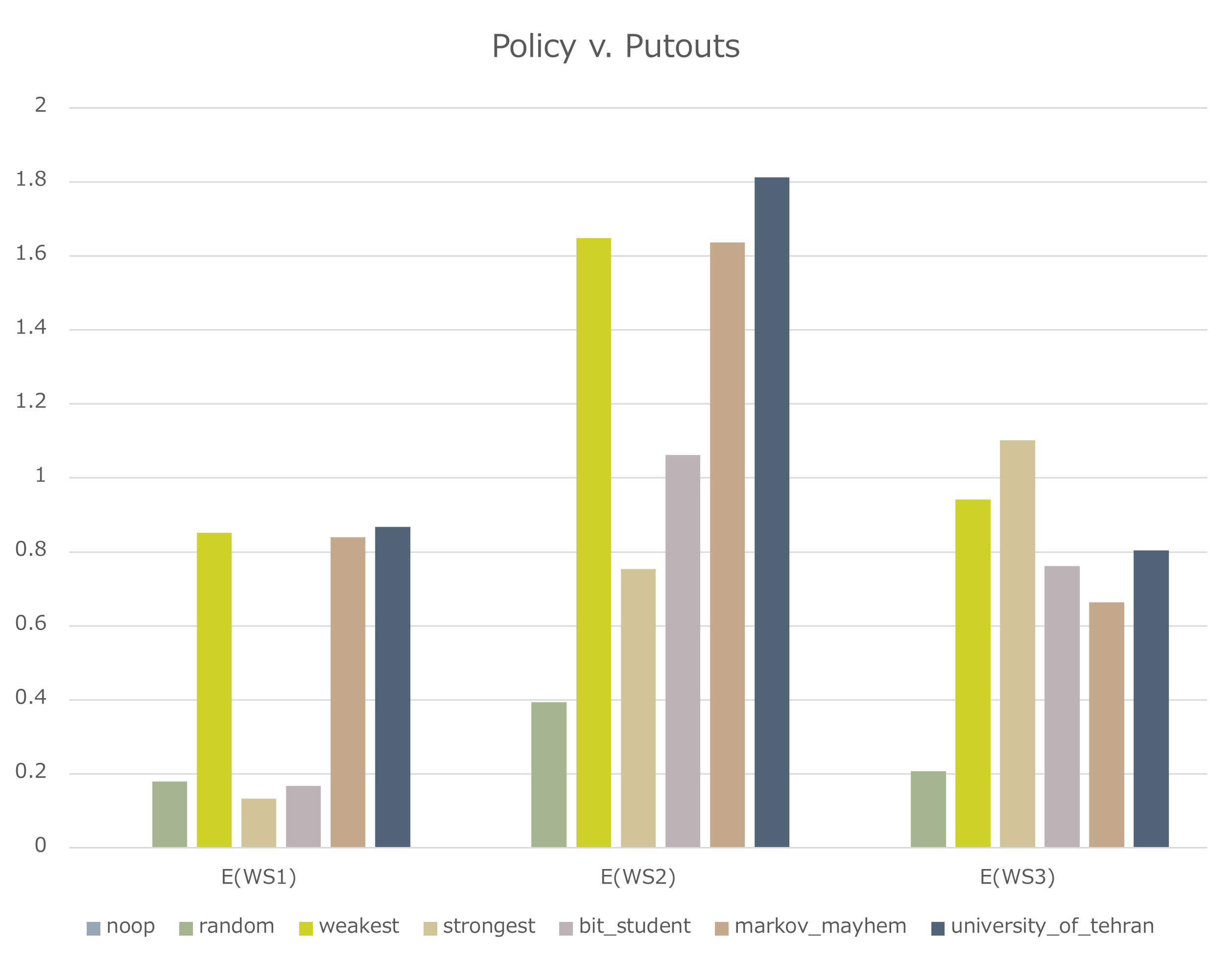}
	\caption{Putouts achieved by each policy in each test configuration}
	\label{fig:wildfire_policy_putouts}
\end{figure}

\begin{figure}[H]
	\centering
	\includegraphics[width=\columnwidth, trim=15 15 15 15, clip]{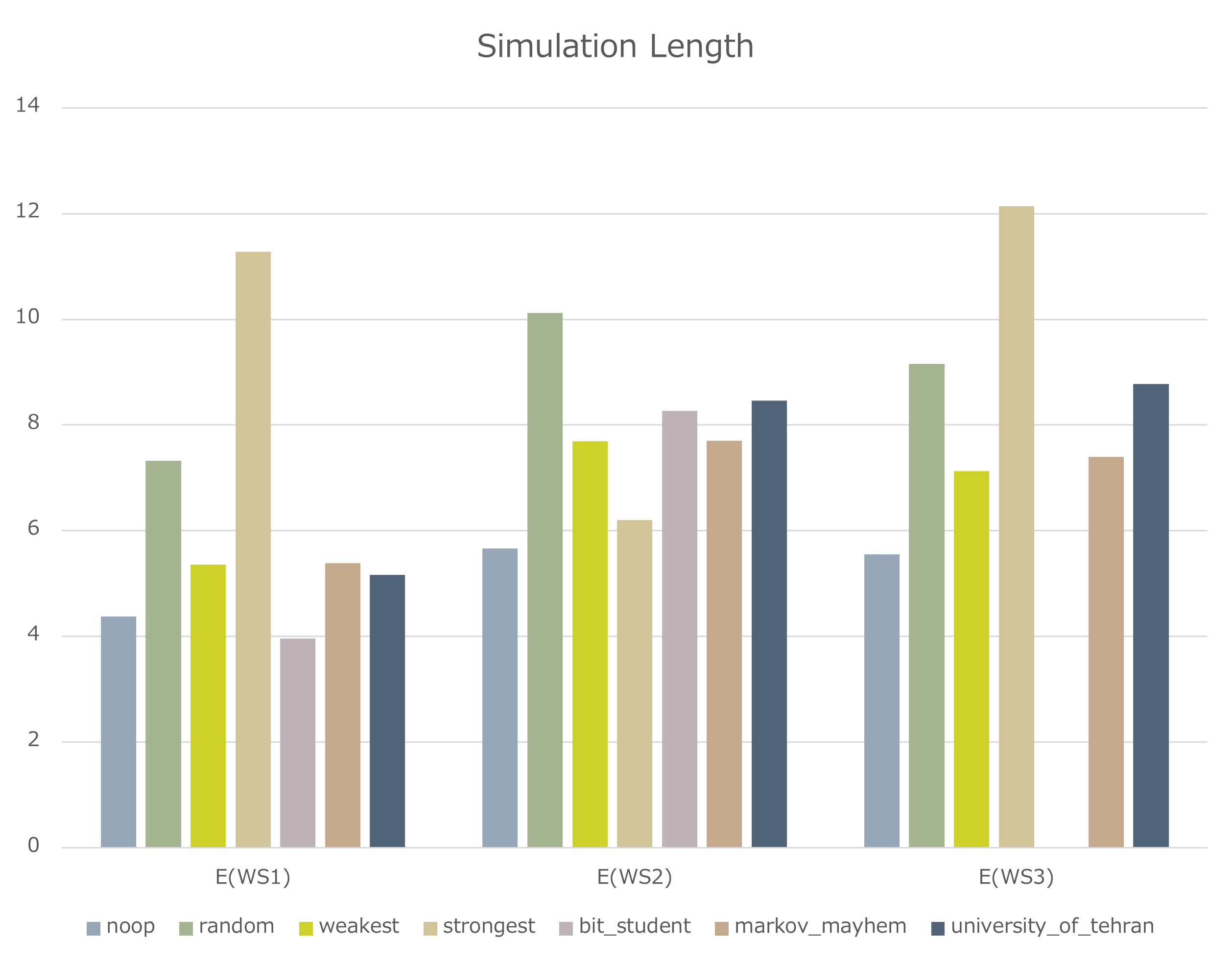}
	\caption{Average length of simulation for each policy in each configuration}
	\label{fig:wildfire_simulation_length}
\end{figure}

Figures \ref{fig:wildfire_policy_burnouts}, \ref{fig:wildfire_policy_putouts}, and \ref{fig:wildfire_simulation_length} provide additional performance metrics in terms of the number of fires that burned out, the number of fires put out, and the average length of simulation, respectively. The number of fires put out is initially highly correlated with the cumulative episodic rewards in WS1, where open conditions less prevalent, however the correlation is no longer found in WS3 where there is significantly more openness in the environment. We observe that neither number of burnouts nor simulation length are correlated with cumulative episodic rewards.

\begin{figure}[H]
	\centering
	\includegraphics[width=\columnwidth, trim=15 15 15 15, clip]{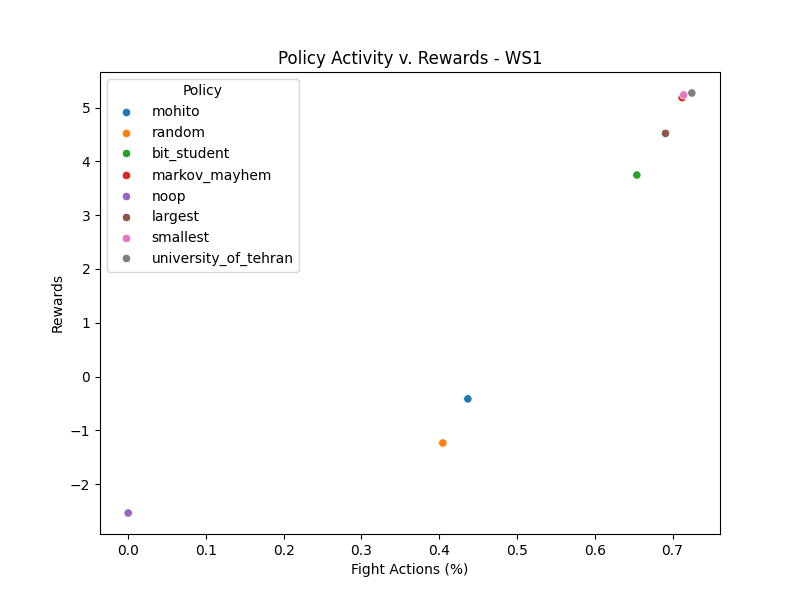}
	\caption{Policy activity (fights) compared against rewards for WS1}
	\label{fig:wildfire_policy_activity_ws1}
\end{figure}

\begin{figure}[H]
	\centering
	\includegraphics[width=\columnwidth, trim=15 15 15 15, clip]{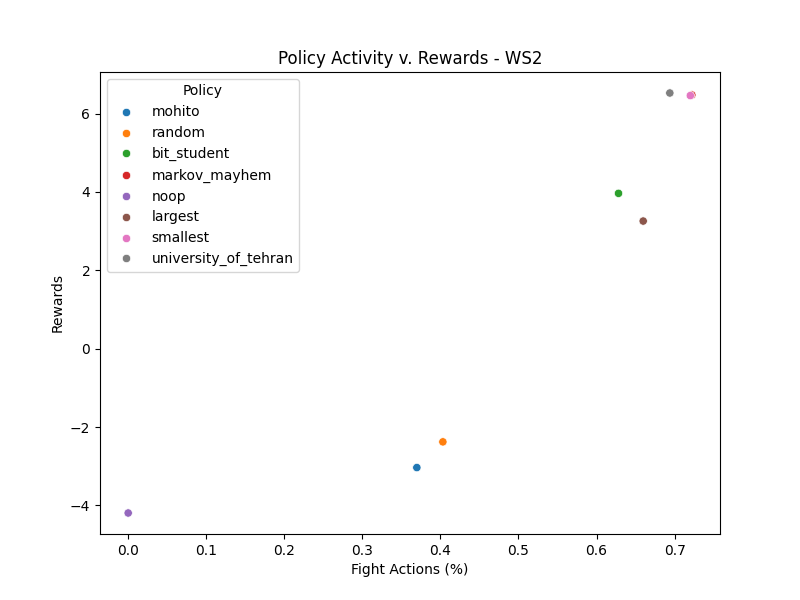}
	\caption{Policy activity (fights) compared against rewards for WS2}
	\label{fig:wildfire_policy_activity_ws2}
\end{figure}

\begin{figure}[H]
	\centering
	\includegraphics[width=\columnwidth, trim=15 15 15 15, clip]{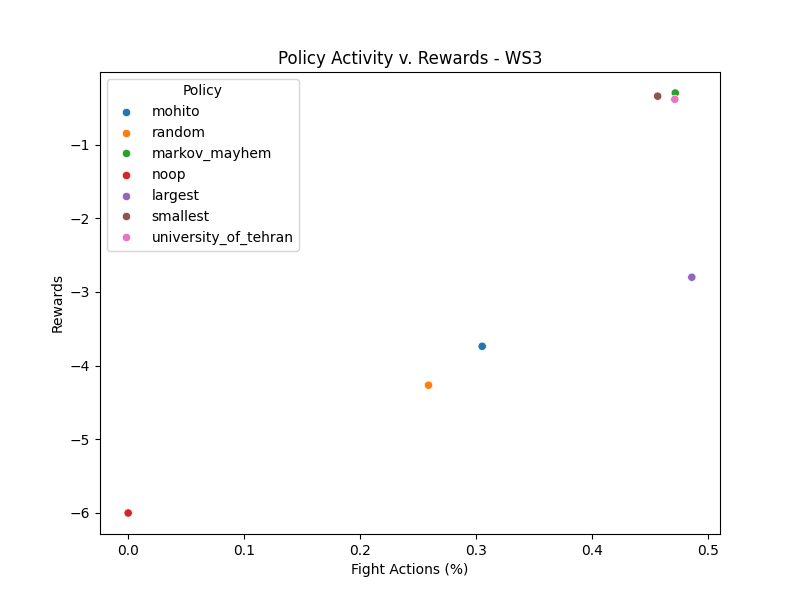}
	\caption{Policy activity (fights) compared against rewards for WS3}
	\label{fig:wildfire_policy_activity_ws3}
\end{figure}

Figures \ref{fig:wildfire_policy_activity_ws1}, \ref{fig:wildfire_policy_activity_ws2}, and \ref{fig:wildfire_policy_activity_ws3} show performance efficiency through the percentage of Fight actions chosen by each approach compared against the rewards each approach receives. In the wildfire domain, we observe that regardless of the level of openness in the environment, more active policies tended to achieve higher episodic rewards. The largest naive approach however spends most of its time fighting fires, however its rewards are significantly lower than Markov Mayhem, University of Tehran, and smallest, indicating that efficiency in resource utilization is still an important strategic factor.

Based on the above results, we recognize the University of Tehran and Markov Mayhem teams as the winners of the 1st MOASEI Competition's wildfire track. We also recognize the BIT Student team as an honorable mention of the 1st MOASEI Competition's wildfire track.

\subsection{Cybersecurity Track Results}

The cybersecurity track featured agent openness, requiring policies to adapt to modifications in the collective strength of not only the their collaborators, but also opposing agents. Rewards are determined by the state of each individual subnetwork within the environment, weighted by the number of outgoing connections to other subnetworks.

Evaluation metrics for this domain included:
\begin{itemize}
	\item\textbf{Average network state over time}: this provides a view into how agents prioritize nodes and gives a summary of their overall strategy for prioritization within the domain. Unstable network states also provide insights into nodes which saw areas of high competition.
	\item\textbf{Action distribution:} Proportions of agent actions (e.g., patching, idling, monitoring / observing, moving) to identify engagement levels and strategic behavior.
\end{itemize}

There were four baseline policies for the cybersecurity domain: 
\begin{enumerate*}[label=(\arabic*)]
  \item\pinoop, which remains idle throughout the simulation;
  \item\pirandom, which uniformly samples valid random actions;
  \item\pipatched, that chooses to patch the least exploited node;
  \item\piexploited, that chooses to patch the most exploited subnetwork node
\end{enumerate*}
We recieved one submission, Zana Cyber, which uses a weighted scoring approach.  

\begin{figure}[H]
	\centering
	\includegraphics[width=\columnwidth, trim=15 15 15 15, clip]{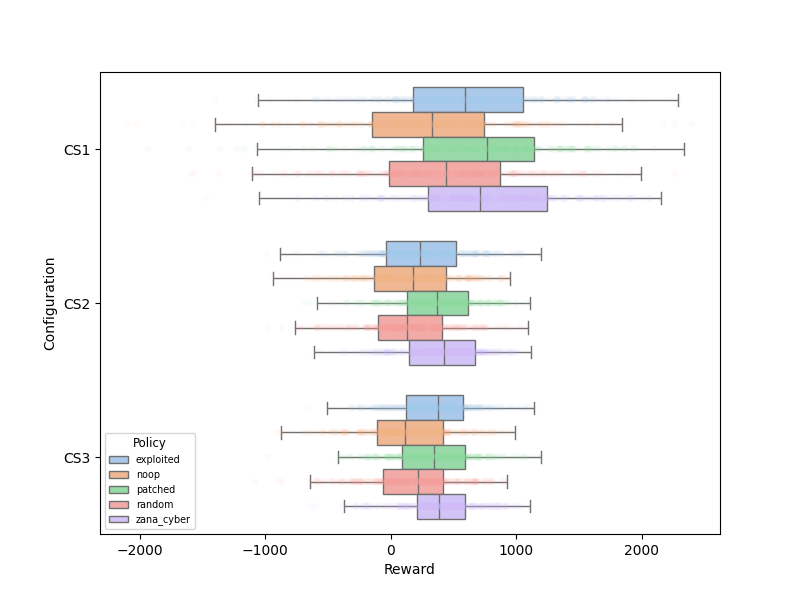}
	\caption{Cybersecurity expected rewards on a per-policy basis}
	\label{fig:cybersecurity_reward_plot}
\end{figure}

\begin{figure}[H]
	\centering
	\includegraphics[width=\columnwidth]{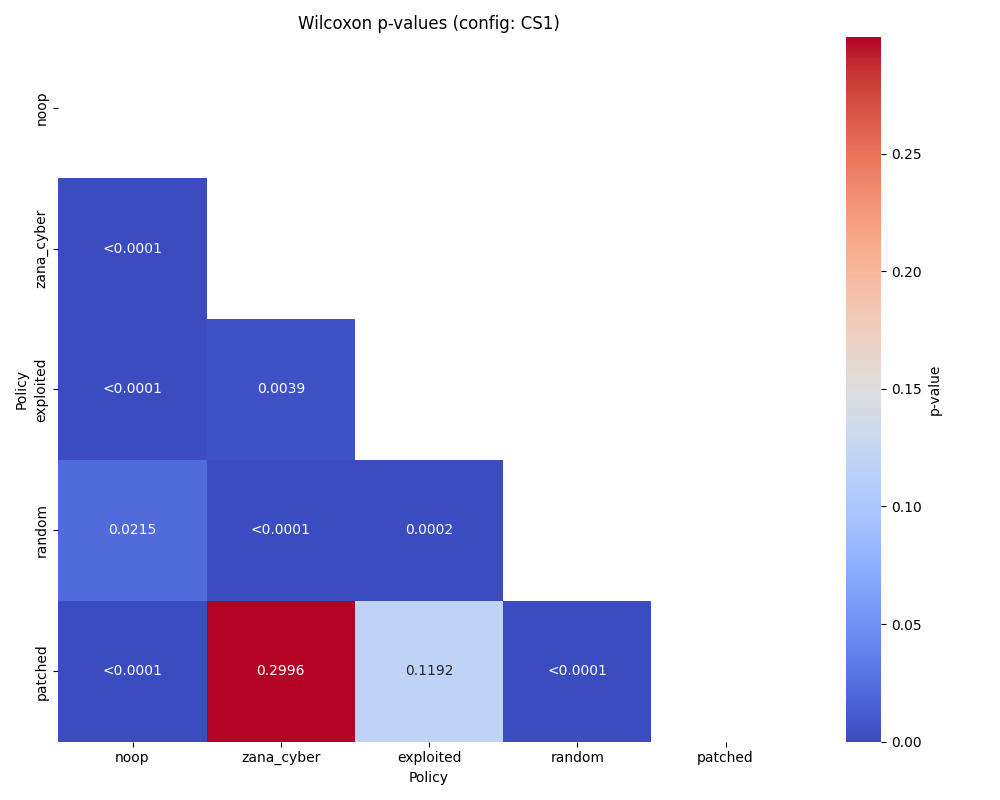}
	\caption{Wilcoxon-signed-rank comparisons for CS1}
	\label{fig:cybersecurity_reward_wilcoxon_cs1}
\end{figure}

\begin{figure}[H]
	\centering
	\includegraphics[width=\columnwidth]{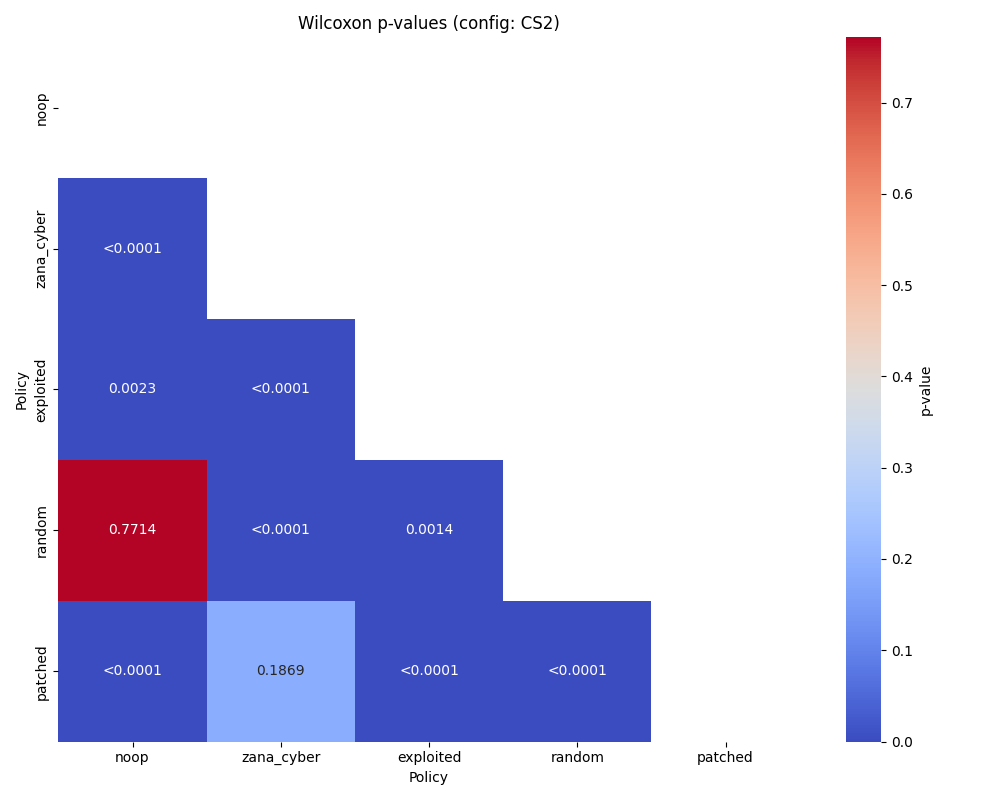}
	\caption{Wilcoxon-signed-rank comparisons for CS2}
	\label{fig:cybersecurity_reward_wilcoxon_cs2}
\end{figure}

\begin{figure}[H]
	\centering
	\includegraphics[width=\columnwidth]{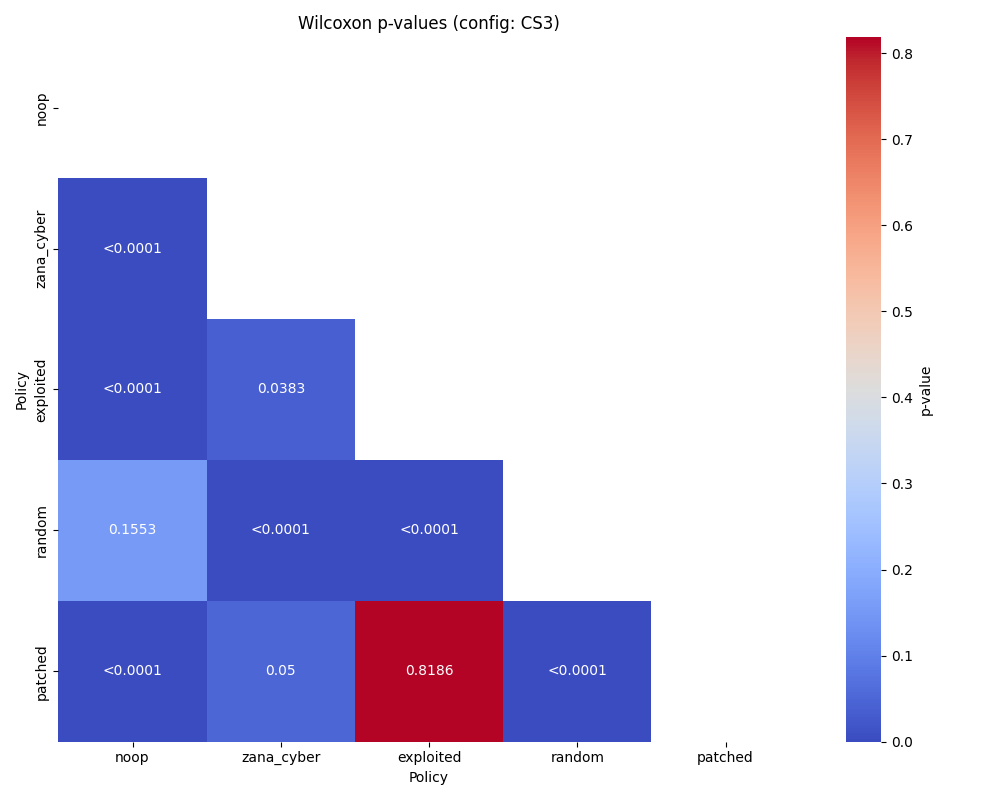}
	\caption{Wilcoxon-signed-rank comparisons for CS3}
	\label{fig:cybersecurity_reward_wilcoxon_cs3}
\end{figure}

Figure~\ref{fig:cybersecurity_reward_plot} shows the expected rewards for baselines and submitted approaches. Figures \ref{fig:cybersecurity_reward_wilcoxon_cs1}, \ref{fig:cybersecurity_reward_wilcoxon_cs2}, and \ref{fig:cybersecurity_reward_wilcoxon_cs3} show Wilcoxon-signed-rank comparisons for the different approaches' mean cumulative episodic rewards to identify statistically signficant differences in rewards. We observe that Zana Cyber achieves the highest episodic rewards across all configurations, however there are no statistically signficiant differences between itself and the patched and exploited approach.

Tables \ref{tab:cybersecurity_policy_move}, \ref{tab:cybersecurity_policy_noop}, \ref{tab:cybersecurity_policy_patch}, and \ref{tab:cybersecurity_policy_monitor} show the proportions of choosing to move to a node, doing nothing, monitoring, and patching, for all approaches, respectively. Here, we observe that Zana Cyber's submission tends towards significantly less patch actions than the other baseline policies, while also performing signficiantly more monitor actions.

\begin{figure}[H]
	\centering
	\includegraphics[width=\columnwidth, trim=5 10 5 10, clip]{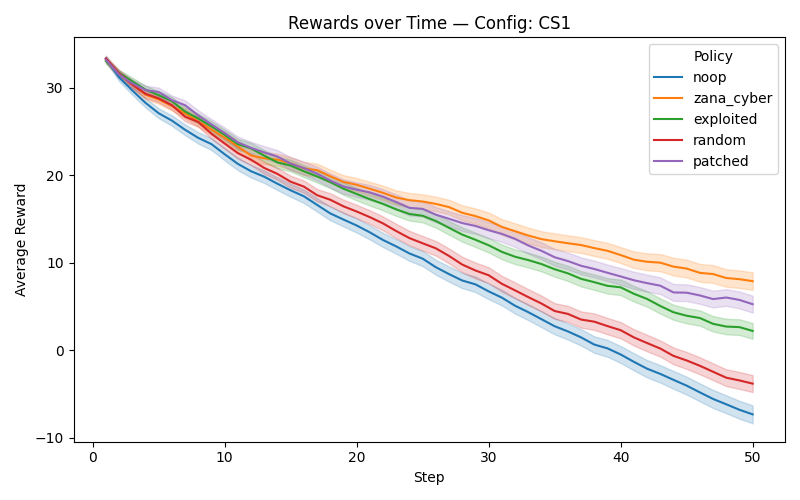}
	\caption{Rewards over time for the cybersecurity domain configuration CS1}
	\label{fig:cybersecurity_reward_over_time_cs1}
\end{figure}

\begin{figure}[H]
	\centering
	\includegraphics[width=\columnwidth, trim=5 10 5 10, clip]{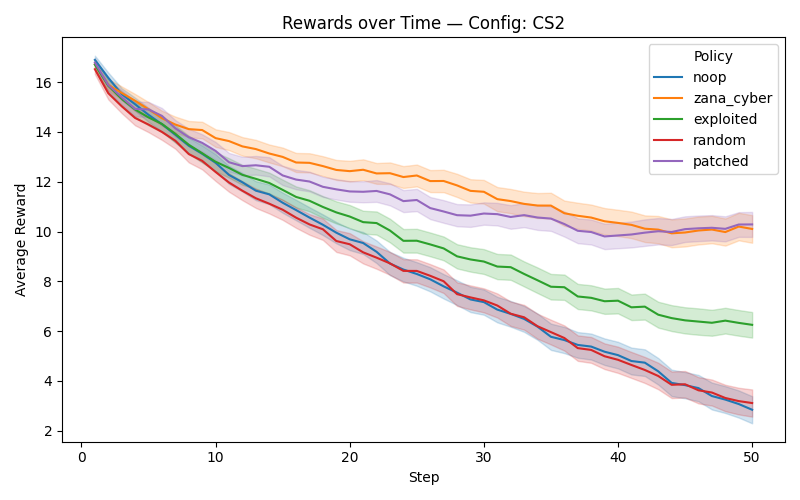}
	\caption{Rewards over time for the cybersecurity domain configuration CS2}
	\label{fig:cybersecurity_reward_over_time_cs2}
\end{figure}

\begin{figure}[H]
	\centering
	\includegraphics[width=\columnwidth, trim=5 10 5 10, clip]{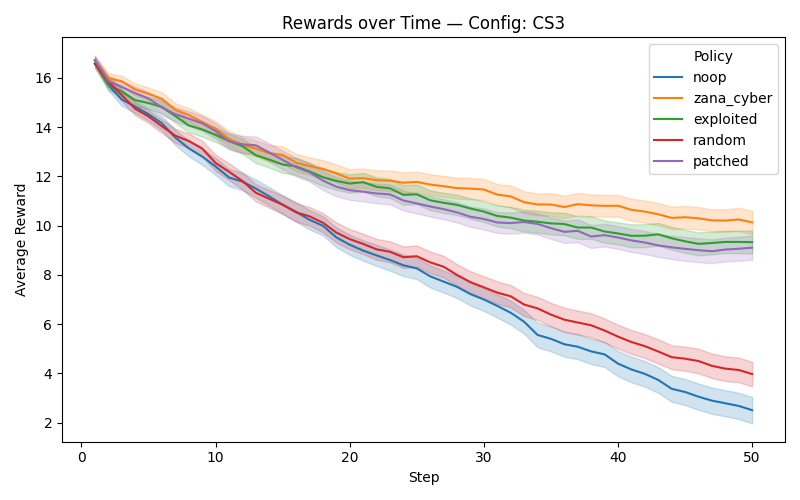}
	\caption{Rewards over time for the cybersecurity domain configuration CS3}
	\label{fig:cybersecurity_reward_over_time_cs3}
\end{figure}

Figures \ref{fig:cybersecurity_reward_over_time_cs1}, \ref{fig:cybersecurity_reward_over_time_cs2}, and \ref{fig:cybersecurity_reward_over_time_cs3} provide additional insights into the performance of each of the approaches over the course of the simulation. Each graph shows the average rewards received by agents at each timestep, which is indicative of the state of the network in its entirety. We note that all policies are progressively losing ground and the subnetwork states are consistently degrading, however the highest performing policies are more effectively resist the efforts of the attacker agents. Zana Cyber's approach in particular shows a much slower downward trend in rewards, indicating more effective resistance by the defender agents.

\begin{figure}[H]
	\centering
	\includegraphics[width=\columnwidth, trim=15 15 15 15, clip]{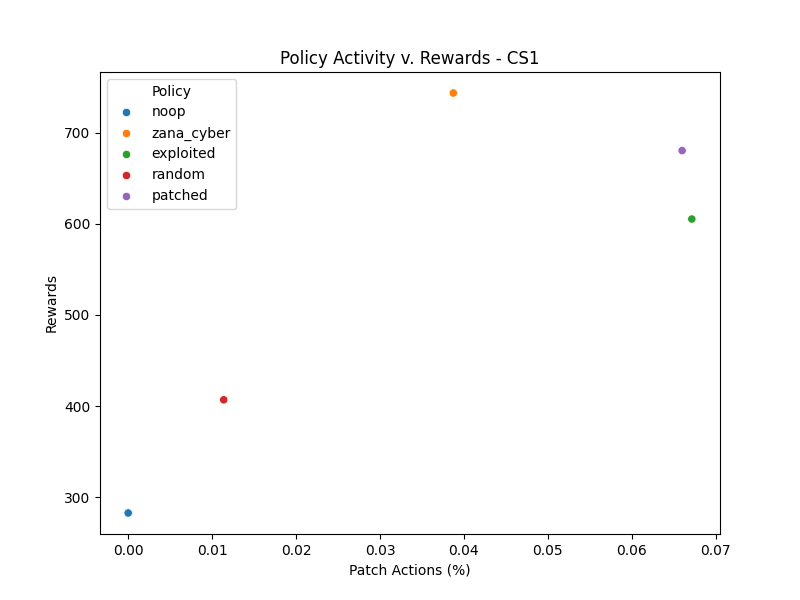}
	\caption{Policy activity (patches) compared against rewards for CS1}
	\label{fig:cybersecurity_policy_activity_cs1}
\end{figure}

\begin{figure}[H]
	\centering
	\includegraphics[width=\columnwidth, trim=15 15 15 15, clip]{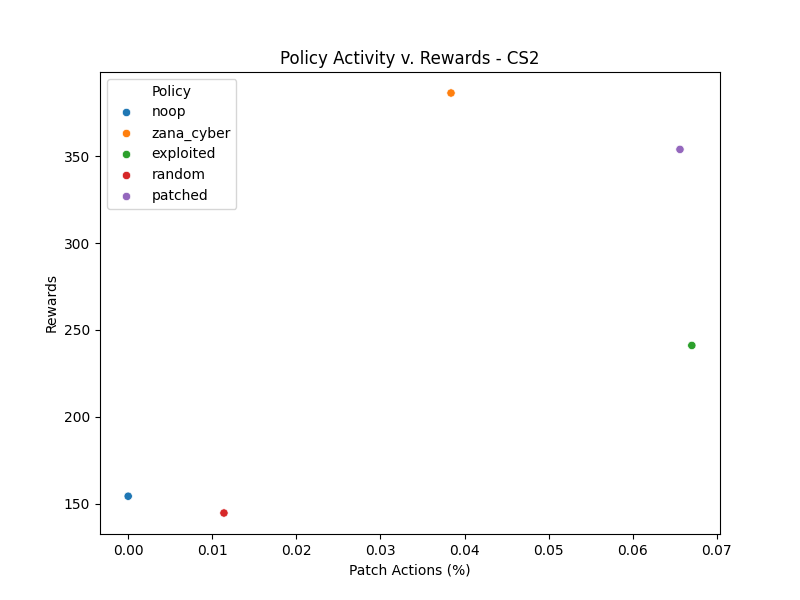}
	\caption{Policy activity (patches) compared against rewards for CS2}
	\label{fig:cybersecurity_policy_activity_cs2}
\end{figure}

\begin{figure}[H]
	\centering
	\includegraphics[width=\columnwidth, trim=15 15 15 15, clip]{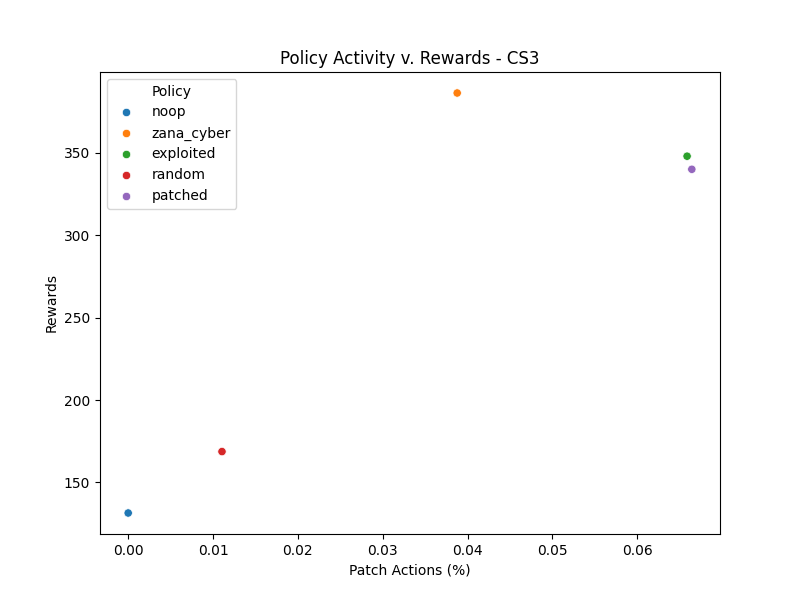}
	\caption{Policy activity (patches) compared against rewards for CS3}
	\label{fig:cybersecurity_policy_activity_cs3}
\end{figure}

Figures \ref{fig:cybersecurity_policy_activity_cs1}, \ref{fig:cybersecurity_policy_activity_cs2}, and \ref{fig:cybersecurity_policy_activity_cs3} show performance efficiency with three scatter plots of the percentage of Patch actions chosen by each approach against the cumulative episodic rewards of each policy, similar to those discussed earlier for the wildfire domain. In this case, a general trend can be observed for the naive baselines: the more agents choose to patch, the more rewards they receive. However, Zana Cyber's approach shows higher episodic rewards when compared to all baselines, while at the same time performing significantly less patch actions. Combined with observations from the policy tables previously indicates that Zana Cyber's submitted policy is more efficient in its application of patch actions by spending more time gathering accurate observations of the environment.

Based on the above results, we recognize the Zana Cyber team as the winner of the 1st MOASEI Competition's cybersecurity track.  

\subsection{Rideshare Track Results}

Unfortunately, there were no submissions to the Rideshare track.  For future MOASEI competitions, we plan to revise this track to make it more accessible and engaging for teams to participate.

\section{Findings from the MOASEI Competition}

The MOASEI Competition served as a large-scale empirical evaluation of multi-agent decision-making under open-agent and open-task conditions, testing agent and task openness across complex, dynamic domains. We discuss key findings about the performances of each of the expected rewards

\subsection{Key Competition Insights}

The competition highlighted a diverse set of solution methods, with a large variety in the classes of approaches for handling open environment. Here we outline the key insights gained from submitted solutions when handling open environments.

\textbf{GNN Methods for Robustness Against Openness}: Teams leveraging relational GNNs demonstrated high adaptability to agent dropout and dynamic tasks. These policies leverage context-aware embeddings included in the rich graph structures, which capture relational structures in each environmental step, enabling better task selection and emergent coordination. Discussions post-competition also highlighted the interpretability benefits provided by the richness of these architectures.

\textbf{CNN Methods for Robustness Against Openness}: CNN-based policies were also shown to match the peak performance of the GNN-based policies. Teams utilizing spatial convolutions in order to facilitate task selection also showed comparable performance in the wildfire domain. These architectures demonstrated a viable lightweight alternative to graph-based methods.

\textbf{Opportunities Utilizing Augmented Loss Functions and Pre-trained Predictors}: Several teams utilized pre-trained predictors that were trained using supervised learning to mitigate the problems posed by partial observability. By pretraining predictors, agents were able to better model the intent of the other agents, which allowed for better coordination within policies. Teams also saw success with non-standard augmented loss functions. These provide models with auxiliary goals that help create a stable gradient during training.

\textbf{Viability of Large Language Models (LLMs) for Handling Open Environments}: One participating team in the MOASEI Competition demonstrated early success using LLMs to iteratively refine a baseline policy through self-prompting and episodic feedback. Rather than relying on traditional end-to-end reinforcement learning pipelines, this team adopts a meta-optimization loop in which an LLM (e.g., GPT-like architecture) generated policy modifications by analyzing past performance, then injected the modified strategy into the environment for further testing. This result illustrates the viability of using LLMs not as direct decision-makers, but as meta-policy optimizers capable of steering policy adaptation over time in response to observed openness. This approach is especially relevant in open environments, where the ability to generalize from sparse or shifting conditions is often more important than maximizing short-term reward. Although the final policy did not outperform the best GNN-based entries in average reward, this brings the possibility of hybrid designs for future research: combining LLM-guided meta-adaptation with graph-based policy architectures may yield systems that both generalize across structural openness and adapt rapidly to unanticipated shifts.

\section{Future Work}

Building on the insights and feedback from MOASEI 2025, we are planning several enhancements for the next iteration of the competition at AAMAS’2026. These enhancements will increase the complexity and realism of each domain, introduce new forms of openness, and provide a more rigorous testbed for evaluating generalization and robustness in multi-agent policies.

In the wildfire domain, we will increase the environment’s spatial scale by expanding the world size and adding more agents and tasks. These changes will place a greater burden on coordination and long-horizon planning, forcing policies to grapple with resource constraints and prioritization in large, dynamic settings. Additionally, the wildfire track will introduce frame openness, dynamic changes to agent capabilities over time. For example, agents may experience equipment degradation or return from downtime with altered skill sets or equipment. This internal variability will require policies to adapt not only to external changes in agent and task presence but also to evolving agent abilities, significantly raising the difficulty and realism of the domain.

The cybersecurity domain will undergo two major upgrades. First, the attacker policies will be enhanced to take full advantage of the observations of the environment, and will have different variations of attacking nodes based on their vulnerability. This creates new opportunities for defenders to incorporate predictive reasoning and anticipation into their policies. Second, the size of the subnetwork and the number of defensive agents will be increased, raising the complexity of the environment and encouraging deeper inter-agent dependencies and cooperative strategies.

We also plan to reintroduce the rideshare domain with a stronger emphasis on direct competition. In this setup, submitted agent policies will directly compete in the same environment for passengers who appear randomly on the map, each with distinct destinations and fares. This structure is expected to encourage policies that are not only efficient in task fulfillment but also robust against adversarial interference from competing submissions.

To further support research continuity and community engagement, we also intend to release a public archive of past submissions and leaderboards. This will help establish long-term benchmarks and facilitate reproducibility and comparison across future editions.

Together, these planned improvements will increase the validity and rigor of our competition to evaluate solutions to open systems. We believe that these modification will allow use to gain deeper insights into how intelligent agents can reason, adapt, and coordinate under diverse and evolving open-system conditions.

\section{Conclusion}
\balance

The inaugural MOASEI Competition thourougly demonstrated the viability of benchmarking AI policies under recognized forms of openness in multi-agent environments. Across three diverse domains--wildfire, cybersecurity, and rideshare--the competition surfaced four substantially different solution approaches, including GNN-based reasoning, CNN-led perception, and LLM-augmented coordination. This diversity highlights the richness of the problem space and affirms the value of MOASEI and the free-range-zoo framework as tools for driving research in open-system AI.

The competition infrastructure proved robust, enabling large-scale evaluation through reproducible pipelines for fixed-seed testing on held-out configurations. Participant engagement was strong, with multiple teams submitting to the wildfire and cybersecurity tracks. Feedback from teams and observers validated both the accessibility of the platform and the novelty of the challenge.

By capturing the nuances of agent and task openness, MOASEI provided a rare testbed for investigating generalization, robustness, and emergent behavior in AI systems with respect to open environments--a key concern for real-world deployment. The success of this initial iteration lays the groundwork for an expanded, recurring benchmark in future years.

\section{Acknowledgements} 
This research was supported, in part, by a collaborative NSF Grant \#IIS-2312657 (to Doshi), \#IIS-2312658 (to Soh), and \#IIS-2312659 (to Eck). Some of the computing occurred on the Holland Computing Center of the University of Nebraska, which receives support from the university's Office of Research and Economic Development and the Nebraska Research Initiative. 

\bibliographystyle{ACM-Reference-Format} 
\bibliography{moasei2025/technical_report}

\begin{table*}[htbp]
  \centering
	\caption{Wildfire Expected Rewards Across Submitted and Baseline Policies}
  \begin{tabular}{lccccc}
    \toprule
		\textbf{Team / Policy}		    & \textbf{WS1}			        & \textbf{WS2}			        & \textbf{WS3}				      & \textbf{Total} 			      \\
    \midrule
		noop							            & $-2.54 \pm 0.05$	        & $-4.20 \pm 0.06$	        & $-6.00 \pm 0.00$		      & $-12.73 \pm 0.12$			    \\
		random						            & $-1.23 \pm 0.15$	        & $-2.38 \pm 0.17$	        & $-4.27 \pm 0.12$		      & $-7.88 \pm 0.44$			    \\
		smallest					            & $5.23 \pm 0.16$		        & $6.46 \pm 0.23$ 	        & $-0.34 \pm 0.18$		      & $11.35 \pm 0.57$			    \\
		largest						            & $4.52 \pm 0.18$		        & $3.26 \pm 0.22$ 	        & $-2.80 \pm 0.17$		      & $4.98 \pm 0.57$			      \\
		bit\_student (LLM)				    & $3.75 \pm 0.09$		        & $3.97 \pm 0.18$ 	        & $crash$					          & $7.71 \pm 0.27$			      \\
		markov\_mayhem (GNN)			    & $5.18 \pm 0.16$		        & $6.48 \pm 0.23$ 	        & $\mathbf{-0.30 \pm 0.22}$	& $11.36 \pm 0.61$			    \\
		university\_of\_tehran (CNN)	& $\mathbf{5.27 \pm 0.16}$	& $\mathbf{6.53 \pm 0.25}$	& $-0.39 \pm 0.21$			    & $\mathbf{11.41 \pm 0.63}$	\\
    \bottomrule
  \end{tabular}
	\label{tab:wildfire_policy_rewards}
\end{table*}

\begin{table*}[htbp]
  \centering
	\caption{Wildfire $a_{fight}$ Action Proportions Across Submitted and Baseline Policies}
  \begin{tabular}{lccccc}
    \toprule
		\textbf{Team / Policy}			  & \textbf{WS1}			    & \textbf{WS2}	  		  & \textbf{WS3}			    & \textbf{Average} 		  \\
    \midrule
		noop							            & $0.00\% \pm 0.00\%$	  & $0.00\% \pm 0.00\%$	  & $0.00\% \pm 0.00\%$	  & $0.00 \% \pm 0.00\%$ 	\\
		random						            & $40.46\% \pm 0.01\%$	& $40.31\% \pm 0.01\%$	& $25.91\% \pm 0.01\%$	& $35.56 \% \pm 0.02\%$	\\
		smallest					            & $71.44\% \pm 0.01\%$	& $72.01\% \pm 0.01\%$	& $45.68\% \pm 0.01\%$	& $63.05 \% \pm 0.03\%$	\\
		largest						            & $48.63\% \pm 0.01\%$	& $69.10\% \pm 0.01\%$	& $65.98\% \pm 0.01\%$	& $61.24 \% \pm 0.02\%$	\\
		bit\_student (LLM)				    & $36.63\% \pm 0.02\%$	& $62.82\% \pm 0.01\%$	& $crash$		    	      & $54.95 \% \pm 0.04\%$	\\
		markov\_mayhem (GNN)			    & $71.24\% \pm 0.01\%$	& $72.21\% \pm 0.01\%$	& $47.20\% \pm 0.01\%$	& $63.55 \% \pm 0.03\%$	\\
		university\_of\_tehran (CNN)	& $72.48\% \pm 0.01\%$	& $69.39\% \pm 0.01\%$	& $47.16\% \pm 0.01\%$	& $63.01 \% \pm 0.03\%$	\\
    \bottomrule
  \end{tabular}
	\label{tab:wildfire_policy_fights}
\end{table*}

\begin{table*}[htbp]
  \centering
	\caption{Wildfire $a_{noop}$ Action Proportions Across Submitted and Baseline Policies}
  \begin{tabular}{lccccc}
    \toprule
		\textbf{Team / Policy}		    & \textbf{WS1}			    & \textbf{WS2}	  		  & \textbf{WS3}			    & \textbf{Average} 		  \\
    \midrule
		noop							            & $100.00\% \pm 0.00\%$	& $100.00\% \pm 0.00\%$	& $100.00\% \pm 0.00\%$	& $100.00 \% \pm 0.00\%$\\
		random						            & $59.54\% \pm 0.01\%$	& $59.69\% \pm 0.01\%$	& $74.00\% \pm 0.01\%$	& $64.44 \% \pm 0.02\%$	\\
		smallest					            & $28.56\% \pm 0.01\%$	& $27.99\% \pm 0.01\%$	& $54.00\% \pm 0.01\%$	& $36.95 \% \pm 0.03\%$	\\
		largest						            & $51.37\% \pm 0.01\%$	& $30.90\% \pm 0.01\%$	& $34.00\% \pm 0.01\%$	& $38.76 \% \pm 0.02\%$	\\
		bit\_student (LLM)			      & $63.37\% \pm 0.02\%$	& $37.18\% \pm 0.01\%$	& $crash$		     	& $45.05 \% \pm 0.04\%$	\\
		markov\_mayhem (GNN)		      & $28.76\% \pm 0.01\%$	& $27.79\% \pm 0.01\%$	& $52.80\% \pm 0.01\%$	& $36.45 \% \pm 0.03\%$	\\
		university\_of\_tehran (CNN)	& $27.52\% \pm 0.01\%$	& $30.61\% \pm 0.01\%$	& $52.84\% \pm 0.01\%$	& $36.99 \% \pm 0.03\%$	\\
    \bottomrule
  \end{tabular}
	\label{tab:wildfire_policy_noop}
\end{table*}

\begin{table*}[htbp]
  \centering
	\caption{Cybersecurity Expected Rewards Across Submitted and Baseline Policies}
  \begin{tabular}{lccccc}
    \toprule
		\textbf{Team / Policy}			    & \textbf{CS1}				  & \textbf{CS2}			  & \textbf{CS3}	   		& \textbf{Total} 		  \\
    \midrule
		noop							              & $282.56 \pm 47.15$		& $154.30 \pm 25.18$	& $131.41 \pm 24.36$	& $568.27 \pm 96.69$	\\
		random						              & $406.83 \pm 43.52$		& $144.64 \pm 24.51$	& $168.70 \pm 22.91$	& $720.16 \pm 90.94$	\\
		patched						              & $680.41 \pm 45.16$		& $353.91 \pm 21.49$	& $340.04 \pm 21.30$	& $1374.36 \pm 87.95$	\\
		exploited					              & $605.30 \pm 40.87$		& $241.08 \pm 24.79$	& $347.99 \pm 20.10$	& $1194.37 \pm 85.75$	\\
		zana\_cyber (weighted scoring)  & $743.65 \pm 42.58$		& $386.37 \pm 22.39$	& $386.34 \pm 20.14$	& $1516.37 \pm 85.11$	\\
    \bottomrule
  \end{tabular}
	\label{tab:cybersecurity_policy_rewards}
\end{table*}

\begin{table*}[htbp]
  \centering
	\caption{Cybersecurity $a_{move}$ Action Proportions Across Submitted and Baseline Policies}
  \begin{tabular}{lccccc}
    \toprule
		\textbf{Team / Policy}				    &	\textbf{CS1}			    &	\textbf{CS2}	  		  &	\textbf{CS3}			    &	\textbf{Average} 		  \\
    \midrule
		noop								              &	$00.00\% \pm 0.00\%$	&	$00.00\% \pm 0.00\%$	&	$00.00\% \pm 0.00\%$	&	$00.00\% \pm 0.00\%$	\\
		random							              &	$24.47\% \pm 0.00\%$	&	$24.57\% \pm 0.00\%$	&	$24.61\% \pm 0.00\%$	&	$24.55\% \pm 0.00\%$	\\
		patched							              &	$22.22\% \pm 0.00\%$	&	$22.24\% \pm 0.00\%$	&	$22.08\% \pm 0.00\%$	&	$22.18\% \pm 0.00\%$	\\
		exploited						              &	$22.12\% \pm 0.00\%$	&	$22.04\% \pm 0.00\%$	&	$21.91\% \pm 0.00\%$	&	$22.02\% \pm 0.00\%$	\\
		zana\_cyber (weighted scoring)		&	$19.89\% \pm 0.00\%$	&	$19.77\% \pm 0.00\%$	&	$19.71\% \pm 0.00\%$	&	$19.79\% \pm 0.00\%$	\\
    \bottomrule
  \end{tabular}
	\label{tab:cybersecurity_policy_move}
\end{table*}

\begin{table*}[htbp]
  \centering
	\caption{Cybersecurity $a_{noop}$ Action Proportions Across Submitted and Baseline Policies}
  \begin{tabular}{lccccc}
    \toprule
		\textbf{Team / Policy}				    &	\textbf{CS1}			    &	\textbf{CS2}	  		  &	\textbf{CS3}			    &	\textbf{Average} 		  \\
    \midrule
		noop								              &	$100.00\% \pm 0.00\%$	&	$100.00\% \pm 0.00\%$	&	$100.00\% \pm 0.00\%$	&	$100.00\% \pm 0.00\%$	\\
		random							              &	$71.46\% \pm 0.00\%$	&	$71.29\% \pm 0.00\%$	&	$71.34\% \pm 0.00\%$	&	$71.36\% \pm 0.00\%$	\\
		patched							              &	$68.36\% \pm 0.00\%$	&	$68.39\% \pm 0.00\%$	&	$68.45\% \pm 0.00\%$	&	$68.40\% \pm 0.00\%$	\\
		exploited						              &	$68.47\% \pm 0.00\%$	&	$68.42\% \pm 0.00\%$	&	$68.53\% \pm 0.00\%$	&	$68.47\% \pm 0.00\%$	\\
		zana\_cyber (weighted scoring)		&	$68.44\% \pm 0.00\%$	&	$68.48\% \pm 0.00\%$	&	$68.57\% \pm 0.00\%$	&	$68.49\% \pm 0.00\%$	\\
    \bottomrule
  \end{tabular}
	\label{tab:cybersecurity_policy_noop}
\end{table*}

\begin{table*}[htbp]
  \centering
  \caption{Cybersecurity $a_{patch}$ Action Proportions Across Submitted and Baseline Policies}
  \begin{tabular}{lccccc}
    \toprule
		\textbf{Team / Policy}			      &	\textbf{CS1}		    &	\textbf{CS2}	  	  &	\textbf{CS3}		    &	\textbf{Average} 	  \\
    \midrule
		noop								              &	$0.00\% \pm 0.00\%$	&	$0.00\% \pm 0.00\%$	&	$0.00\% \pm 0.00\%$	&	$0.00\% \pm 0.00\%$	\\
		random							              &	$1.14\% \pm 0.00\%$	&	$1.14\% \pm 0.00\%$	&	$1.11\% \pm 0.00\%$	&	$1.13\% \pm 0.00\%$	\\
		patched							              &	$6.60\% \pm 0.00\%$	&	$6.56\% \pm 0.00\%$	&	$6.64\% \pm 0.00\%$	&	$6.60\% \pm 0.00\%$	\\
		exploited						              &	$6.59\% \pm 0.00\%$	&	$6.70\% \pm 0.00\%$	&	$6.72\% \pm 0.00\%$	&	$6.67\% \pm 0.00\%$	\\
		zana\_cyber (weighted scoring)		&	$3.87\% \pm 0.00\%$	&	$3.84\% \pm 0.00\%$	&	$3.88\% \pm 0.00\%$	&	$3.86\% \pm 0.00\%$	\\
    \bottomrule
  \end{tabular}
	\label{tab:cybersecurity_policy_patch}
\end{table*}

\begin{table*}[htbp]
  \centering
	\caption{Cybersecurity $a_{monitor}$ Action Proportions Across Submitted and Baseline Policies}
  \begin{tabular}{lccccc}
    \toprule
		\textbf{Team / Policy}				    &	\textbf{CS1}		    &	\textbf{CS2}	  	  &	\textbf{CS3}		    &	\textbf{Average} 	  \\
    \midrule
		noop								              &	$0.00\% \pm 0.00\%$	&	$0.00\% \pm 0.00\%$	&	$0.00\% \pm 0.00\%$	&	$0.00\% \pm 0.00\%$	\\
		random							              &	$2.94\% \pm 0.00\%$	&	$3.01\% \pm 0.00\%$	&	$2.95\% \pm 0.00\%$	&	$2.96\% \pm 0.00\%$	\\
		patched							              &	$2.82\% \pm 0.00\%$	&	$2.81\% \pm 0.00\%$	&	$2.82\% \pm 0.00\%$	&	$2.82\% \pm 0.00\%$	\\
		exploited						              &	$2.82\% \pm 0.00\%$	&	$2.84\% \pm 0.00\%$	&	$2.85\% \pm 0.00\%$	&	$2.84\% \pm 0.00\%$	\\
		zana\_cyber (weighted scoring)    &	$7.80\% \pm 0.00\%$	&	$7.91\% \pm 0.00\%$	&	$7.85\% \pm 0.00\%$	&	$7.85\% \pm 0.00\%$	\\
  \bottomrule
  \end{tabular}
	\label{tab:cybersecurity_policy_monitor}
\end{table*}

\FloatBarrier
\clearpage

\FloatBarrier
\clearpage

\end{document}